\documentstyle[12pt]{article} 
\begin{document} 
 
\def\beq{\begin{equation}} 
\def\eeq{\end{equation}} 
\def\bce{\begin{center}} 
\def\ece{\end{center}} 
\def\bea{\begin{eqnarray}} 
\def\eea{\end{eqnarray}} 
\def\ben{\begin{enumerate}} 
\def\een{\end{enumerate}} 
\def\ul{\underline} 
\def\ni{\noindent} 
\def\nn{\nonumber} 
\def\bs{\bigskip} 
\def\ms{\medskip} 
\def\wt{\widetilde} 
\def\wh{\widehat} 
\def\brr{\begin{array}} 
\def\err{\end{array}} 
\def\dsp{\displaystyle} 
\def\D{{D \hskip -3mm /\,}} 
\def\e{{\rm e}} 
 
\begin{flushright} 
IEEC 98-81 \\ 
NDA-FP-52 \\ 
hep-th/9901026 \\ 
\end{flushright} 
 
\vfill 
 
\begin{center} 
 
{\LARGE \bf Possible quantum instability of primordial black holes}

\vspace{6mm}

{\sc E. Elizalde}$^{a,b,}$\footnote{E-mail: 
 elizalde@ieec.fcr.es, eli@zeta.ecm.ub.es}, 
{\sc S. Nojiri}$^{c,}$\footnote{E-mail:nojiri@cc.nda.ac.jp}  and   
{\sc S.D. Odintsov}$^{a,d,}$\footnote{E-mail: odintsov@tspi.tomsk.su 
} \\ 
\mbox{} \\ 
$^a${\it Consejo Superior de Investigaciones Cient\'{\i}ficas (CSIC),\\ 
Institut d'Estudis Espacials de Catalunya (IEEC), \\ 
Edifici Nexus 201, Gran Capit\`a 2-4, 08034 Barcelona, Spain }\\  
$^b${\it Departament ECM and IFAE, Facultat de F\'{\i}sica, \\ 
Universitat de Barcelona, Diagonal 647, 
08028 Barcelona, Spain} \\ 
$^c${\it Department of Mathematics and Physics, National Defence Academy,\\ 
Hashirimizu Yokosuka 239, Japan}\\ 
$^d${\it  Departament of Mathematics and Physics,} \\ 
{\it Tomsk Pedagogical 
University, 634041, Tomsk, Russia}\\ 
 
\vspace{4mm} 
 
{\bf Abstract} 
 
\end{center} 
 
Evidence for the possible existence of a quantum process opposite to the  
famous Hawking radiation (evaporation) of black holes is presented. 
This new phenomenon could be very relevant in the case of exotic 
multiple horizon Nariai black holes and in the context of common 
grand unified theories. This is clearly manifested in the case of 
the SO(10) GUT, that is here investigated in detail. The remarkable result  
is obtained,  that anti-evaporation can occur there only in the SUSY version
of the theory.  
It is thus concluded that the existence of primordial black holes in the
 present Universe might be considered as an evidence for supersymmetry. 
 
\vfill 
 
\noindent {\it PACS:}  04.70.Dy, 04.60.-m, 04.62.+v

\newpage 
 
 
In his celebrated paper, Ref. \cite{haw1}, published over twenty 
years ago, Steven Hawking made one of the most striking discoveries of the 
history of black hole physics: the possibility for black holes to evaporate, 
as a result of particle creation. This effect ---which is now called 
the Hawking radiation process--- produced a deep impact in our  
understanding of quantum gravity. Nowadays it has become to be considered  
as a classical test of the theory. 
 
What conferes this effect its great importance is the observation that  
Hawking radiation is a  universal phenomenon. One of its most dramatic 
consequences being the fact that all primordial black holes should 
evaporate in the process of the early universe evolution. 
 
There exists, however, an exotic class of black holes (see Ref. \cite{haw2} 
for a review) which posses a multiple horizon and for which the  
{\it opposite} effect might occur. We have investigated this possibility  
in detail. Consider a nearly degenerate Schwarzschild-de Sitter 
black hole (the so-called Nariai black hole \cite{nari1} in the  
specialized  literature). The Schwarzschild-de Sitter 
black hole represents the neutral, static, spherically symmetric  
solution of Einstein's theory with a cosmological constant.  
The corresponding metric looks as follows 
\beq  
ds^2= -V(\tau )dt^2 + V^{-1}(\tau )d\tau^2+\tau^2 d\Omega^2, 
\qquad
\label{ii} 
V(\tau ) = 1-\frac{2\mu}{\tau} - \frac{\Lambda}{3} \tau^3, 
\eeq 
where $\mu$ is the mass and $\tau$ the radius of the black hole, $d\Omega^2$  
 the metric corresponding to a unit two-sphere,  
and $\Lambda$ is the cosmological constant. It can be easily checked  
that the equation  
\beq 
V(\tau )=0, 
\eeq 
has two positive roots, $\tau_c$ and $\tau_b$ (we set $\tau_c >\tau_b$). 
Here,  
$\tau_c$ and $\tau_b$ have the meaning of a cosmological and a black-hole  
horizon radius, respectively. In the degenerate case (which corresponds to  
a black hole of maximal mass) both radii coincide and the black hole is  
in thermal equilibrium. It is called a Nariai black hole. 
 
There are in fact two opposite sources contributing to this equilibrium,  
namely a radiation flux coming from the cosmological horizon and the Hawking 
evaporation originated at the black hole horizon.  
It is plausible that such a state might be unstable, since it could 
be affected by small perturbations of the geometry. This is what happens,  
indeed. 
 
It was demonstrated in a recent paper by R. Bousso and S.W. Hawking, 
Ref. \cite{bh1}, that a nearly maximal (or nearly degenerated) Nariai 
black hole may not only evaporate ---as shown in Ref. \cite{haw1}--- 
but also anti-evaporate \cite{bh1}. In other words, Nariai black holes  
actually develop two perturbative modes: an evaporating one and an  
anti-evaporating one. The mathematical realization of this quantum process, 
carried out in Ref.  \cite{bh1}, is based on the generalized CGHS model 
of two-dimensional dilaton quantum gravity \cite{cghs1}, with the dilaton 
coupled scalars \cite{e} coming from the so-called minimal  
four-dimensional scalars 
in the process of spherical reduction (for a basic introduction of these  
concepts the reader is addressed to Ref. \cite{bos1}).  
 
The possibility of evaporation or anti-evaporation (i.e., increase in size) 
of a black hole of this kind is certainly connected with the initial  
conditions chosen for the perturbations. In the model of Ref. \cite{bh1}, 
use of the commonly employed Hartle-Hawking no-boundary conditions \cite{hh1} 
shows that such black holes will most likely evaporate. We are thus left  
with the original situation, in this case. 
 
The exotic Nariai black holes are not asymptotically flat. They  
will never appear   in the process of star collapse. Nevertheless, they  
could actually be present in the early inflationary universe, through  
any of the following processes: 
\ben 
\item pair creation of primordial black holes during inflation \cite{bh2}; 
\item quantum generation due to quantum fluctuations of matter fields 
\cite{bno1}. 
\een 
In any case, according to the model in Ref.  \cite{bh1}, primordial  
multiple horizon black holes should quickly evaporate, and it is very  
unlikely that they could be detected in the present universe. 
 
However, a different model for anti-evaporation of black holes has  
been proposed recently, Ref. \cite{no1}, in which these difficulties 
may be overcome. In that model, the quantum effects of the  
conformally invariant matter have been taken into account. What is more  
interesting, this theory allows for the possibility of including not 
only scalar fields, but also fermionic and vector fields (typical of all  
grand unified theories, GUTs), whose classical action is  
\beq 
\label{OI} 
S=\int d^4x \sqrt{-g_{(4)}}\left\{{1 \over 2}\sum_{i=1}^N  
\left(g_{(4)}^{\alpha\beta}\partial_\alpha\chi_i 
\partial_\beta\chi_i + {1 \over 6}R^{(4)}\chi_i^2 \right) 
-{1 \over 4}\sum_{j=1}^{N_1}F_{j\,\mu\nu}F_j^{\mu\nu} 
+\sum_{k=1}^{N_{1/2}}\bar\psi_k\D\psi_k \right\}, 
\eeq 
and its quantum correction $\Gamma$ is the sum of the conformal  
anomaly induced action $W$ and the action $\Gamma'$ given by the  
Schwinger-de Witt type expansion: 
\bea 
\label{OVIII} 
\Gamma&=&W+\Gamma', \nn \\ 
W&=&b\int d^4x \sqrt{-g} F\sigma  
+b'\int d^4x \sqrt{-g} \Bigl\{\sigma\left[ 
2\Box^2 + 4 R^{\mu\nu}\nabla_\mu\nabla_\nu \right. \nn \\ 
&& \left. - {4 \over 3}R\Box + {2 \over 3}(\nabla^\mu R)\nabla_\mu  
\right]\sigma + \left(G-{2 \over 3}\Box R\right)\sigma \Bigr\} \nn \\ 
&& -{b + b' \over 18} 
\int d^4x \sqrt{-g}\left[R - 6 \Box \sigma  
- 6(\nabla \sigma)(\nabla \sigma) \right]^2, \nn \\ 
\Gamma'&=&\int d^4x  
\sqrt{-g}\left\{\left[{b}F+  {b'}G 
+ { 2b\over 3}\Box R\right]\ln { R \over \mu^2}\right\} 
+{\cal O}(R^3). 
\eea 
Here $b={(N +6N_{1/2}+12N_1)\over 120(4\pi)^2}$,  
$b'=-{(N+11N_{1/2}+62N_1) \over 360(4\pi)^2}$, $F$ 
is the square of the Weyl tensor, $G$ the Gauss-Bonnet invariant,  
$\mu$ is a mass-dimensional constant parameter, and we choose the metric  
as $ds^2=\e^{-2\sigma}g_{\mu\nu}dx^\mu dx^\nu$. 
Specially when we assume the solution to be spherically symmetric,  
the metric has the form of $ds^2=\e^{-2\sigma} 
\left(\sum_{\alpha,\beta=0,1}g_{\alpha\beta}dx^\alpha dx^\beta  
+ r_0^2 d\Omega^2\right)$, where $d\Omega^2$ is the metric on the unit 
two-sphere. 
The parameter $r_0$ is introduced by hand. Note that the Schwinger-de Witt  
expansion is the corresponding power expansion with respect to the curvature. 
Having introduced 
the parameter $r_0$, the scalar curvature given by the metric $g_{\mu\nu}$  
is of the order of $1/r_0^2$ if there is no singularity, 
as in the Nariai black  
hole. Therefore, if we choose $r_0$ to be large, the Schwinger-de Witt type  
expansion becomes exact. 
 
By solving the quantum effective equations of motion derived from (\ref{OI})  
and (\ref{OVIII}), we can find the quantum analogue of the Nariai black 
hole,  
which has constant scalar curvature $R=R_0$ and radius  
$\e^\sigma=\e^{\sigma_0}$:  
\bea 
\label{solrl} 
R_0&=&\left\{2+\left( \ln (\mu r_0) \right)^{-1} 
\left({2b+3b' \over b}+{9 \over 512\pi^2 b G\Lambda}\right) \right\}  
+ {\cal O}\left( \left( \ln (\mu r_0) \right)^{-1}\right), \nn \\ 
\sigma_0&=&- \ln (\mu r_0) + {1 \over 2}\ln \left( 
{3\mu^2 \over 2\Lambda}\right) \nn \\ && +\left( \ln (\mu r_0) \right)^{-1} 
{\mu^2 \over 8\Lambda}\left({2b+3b' \over b}+{9 \over 512\pi^2 b G\Lambda} 
\right) + {\cal O}\left( \left( \ln (\mu r_0) \right)^{-1}\right).
\eea 
Here we assume $r_0$ to be large. Furthermore, we can find the  
perturbation around the solution. It is given by an eigenfunction of  
the Laplacian in the two-dimensional hyperboloid, which corresponds to the  
subspace in the Nariai balck hole, given by radial and time coordinates.  
The fate of the perturbed black hole is governed by the eigenvalue of the  
Laplacian and we find that  anti-evaporation can occur only if  
\beq 
\label{Ns} 
2N+7N_{1/2}>26 N_1 . 
\eeq 
 
Owing to the fact that the equations of motion in the model by 
Bousso and Hawking contain 
only second order derivatives, anti-evaporation is excluded there by the  
no-boundary condition of Hartle and Hawking. The quantum effective  
equations of motion given by (\ref{OI}) and (\ref{OVIII}), however,  
contain  fourth order derivatives and there the anti-evaporation 
phenomenon can be 
consistent with the  no-boundary condition.  
 
As an example take, for instance, the usual SO(10) GUT, which would be a  
typical model in the early universe. First, we consider  
the non-supersymmetric model of \cite{ceg}, with $16\times 
3$ (generation) fermions and $16+120+(10$ or $126)=136$ or $262$ Higgs  
scalars (the numbers are the dimensions of the representations).  
Therefore we obtain $2N+7N_{1/2}=608$ or $860$, and  
 the adjoint vector fields $(N_1=45)$ give a contribution  
of $26N_1=1170$. Thus, we find that Eq. (\ref{Ns}) cannot be  
satisfied. The situation is, however, drastically changed when we  
consider the supersymmetric model. In the naive supersymmetric 
extension of the above model, we have contributions from squarks, 
Higgsinos and gauginos. Including them, we find that  
$2N+7N_{1/2}=1971$ or $3105$, and Eq. (\ref{Ns}) is then certainly 
satisfied, since  
the contribution from the vector fields does not change from the one of the 
non-supersymmetric model. The above structure is not modified either in 
the  various extensions of the SO(10) model that have been considered 
more recently. That is, in all these cases 
for the non-supersymmetric model 
Eq. (\ref{Ns}) is {\it not} satisfied, since the  
contribution from the vector fields dominates, but for the 
supersymmetric models the contribution from the Higgsino  
in the large dimensional representations dominates and  
Eq. (\ref{Ns}) {\it is} fulfilled. Then, under the hypothesis that the  
no-boundary initial condition applies, at least some primordial  
multiple horizon black holes are going to anti-evaporate. 
 
In other words, primordial black holes might exist for a much longer period 
 (if we consider some reasonable SUSY GUT as a realistic theory) 
than it had been expected \cite{haw1}. Primordial black holes would 
have grown,  
for some time, until other effects could have stopped the process. Some could  
even have survived and be present in our  universe. Then, 
if these primordial black holes were 
observed, they would constitute an indirect evidence of supersymmetry.  
The rate $P$ of  pair creation of  black holes has appeared in \cite{bh2}: 
\beq 
\label{rate} 
P=\exp\left(-{\pi \over G \Lambda_{eff}}\right). 
\eeq 
In the Euclidean path integral, the weight of the probability is given  
---in the semi-classical approximation--- by substituting the 
 classical black 
hole solution $g_{\mu\nu}^{\rm classical}$,  
after Wick rotating $t\rightarrow i\tau$ ($\tau$ has a period of  
$\sqrt{1 \over \Lambda}{\pi \over 2}$ for Nariai black hole and  
$\sqrt{3 \over \Lambda}{\pi \over 2}$ for anti-de Sitter space),  
into the action $S$ in (\ref{OI})  
and exponentiating the action $\e^{-S(g_{\mu\nu}^{\rm classical})}$.  
Eq. (\ref{rate}) is evaluated from the ratio of the weights for the classical  
Nariai space and anti-de Sitter space:  
$P=\e^{-S(g_{\mu\nu}^{\rm classical\ Nariai}) 
+S(g_{\mu\nu}^{\rm classical\ anti-de\ Sitter})} 
=\exp\left(-{\pi \over G \Lambda}\right)$. 
In the inflational universe, the effective cosmological constant  
$\Lambda_{eff}$ in Eq. (\ref{rate}) is typically given by the square  
of the GUT scale, $10^{16}$ GeV; then the exponent  
${\pi \over G \Lambda_{eff}}$ in Eq. (\ref{rate})  
is of the order of $10^6$ and  pair production would  
be suppressed. But the magnitude of the exponent depends on the model  
of inflation used. If our universe were pair created as in the original 
work by Hartle and Hawking \cite{hh1}, in the very early universe the 
energy density could be of the  
order of the Planck scale and  then
$\Lambda_{eff}$ would be of the order of unity. In such a situation, the  
pair-creation of the Nariai black hole would  not be suppressed.  
The important consequence being then, 
 that the pair created black hole would 
{\it not} evaporate and could survive in the present universe.   

In summary, we conclude that adopting the  SO(10) model (in any of its several
versions) and under the hypothesis that the  
no-boundary initial condition applies, we could obtain a proof of 
supersymmetry by finding evidence of the existence of anti-evaporating
black holes. One may argue that, even if they  existed, the probability of
finding just one black hole of this kind would be very low. However, 
in view of the extreme importance of supersymmetry for all string, brane
 and conformal theories, and the like, and in view also of the extreme 
difficulty in proving, by {\it any} other means, that such a strongly
 broken symmetry as SUSY is
 in fact a symmetry of nature (and not merely a beautiful idea),
conferes a relevant status to the results we have here obtained.

Again, we concede that the process of anti-evaporation might in fact
 be rather exotic and only limited to the very early times of  
the inflationary universe.  In the own words of Steven Hawking: {\it I regard 
anti-evaporation as a pathology}. Notwithstandingthat, the model above  
has appeared as a brand new mathematical  
solution of plausible quantum gravity equations under 
the conditions that are usually assumed to be the natural ones during the  
evolution of the early universe. It would not be wise to discard such  
solution ---and the physical scenario it gives rise to--- before  
an observational quest is consistently pursued. Moreover,  
anti-evaporation might also lie on the basis of other cosmological  
effects (like inflation) in the very early universe. This is  
 presently being investigated by 
the authors and the results will be reported elsewhere.

\vspace{1cm} 
 
One of us (S.D.O.) would like to thank Prof. S.W. Hawking for the kind  
hospitality at DAMTP and for very interesting discussions.  
 This investigation has been  supported by 
 DGICYT (Spain), project 
PB96-0925. 

\end{document}